\documentclass[pra,aps,10pt,superscriptaddress,twocolumn,floatfix]{revtex4-1}
\usepackage{graphicx,color}
\usepackage[nice]{nicefrac}							
\usepackage{amsmath,amssymb,bm}

\usepackage[plainpages=false,pdfpagelabels,colorlinks=true,linkcolor=red,urlcolor=blue,citecolor=blue,pdftitle={},pdfauthor={},pdfdisplaydoctitle=true,pdfduplex=DuplexFlipLongEdge]{hyperref}

\definecolor{darkred}{rgb}{0.90,0.2,0.2}
\definecolor{darkgreen}{rgb}{0,0.60,.2}
\definecolor{darkblue}{rgb}{0.1,0.3,1}
\definecolor{grey}{cmyk}{0,0,0,0.25}
\definecolor{orange}{cmyk}{0,0.6,0.8,0}

\begin{document}

\title{Ergodicity Breaking Transition in Finite Disordered Spin Chains}

\author{Jan \v Suntajs}
\affiliation{Department of Theoretical Physics, J. Stefan Institute, SI-1000 Ljubljana, Slovenia}
\author{Janez Bon\v ca}
\affiliation{Department of Physics, Faculty of Mathematics and Physics, University of Ljubljana, SI-1000 Ljubljana, Slovenia}
\affiliation{Department of Theoretical Physics, J. Stefan Institute, SI-1000 Ljubljana, Slovenia}
\author{Toma\v z Prosen}
\affiliation{Department of Physics, Faculty of Mathematics and Physics, University of Ljubljana, SI-1000 Ljubljana, Slovenia}
\author{Lev Vidmar}
\affiliation{Department of Theoretical Physics, J. Stefan Institute, SI-1000 Ljubljana, Slovenia}
\affiliation{Department of Physics, Faculty of Mathematics and Physics, University of Ljubljana, SI-1000 Ljubljana, Slovenia}


\begin{abstract}
We study disorder-induced ergodicity breaking transition in high-energy eigenstates of interacting spin-1/2 chains.
Using exact diagonalization we introduce a cost function approach to quantitatively compare different scenarios for the eigenstate transition.
We study ergodicity indicators such as the eigenstate entanglement entropy and the spectral level spacing ratio, and we consistently find that an (infinite-order) Kosterlitz-Thouless transition yields a lower cost function when compared to a finite-order transition.
Interestingly, we observe that the transition point in finite systems exhibits nearly thermal properties, i.e., ergodicity indicators at the transition are close to the random matrix theory predictions.
\end{abstract}

\maketitle

{\it Introduction.}
Generic quantum many-body systems are expected to be quantum ergodic, implying that long-time averages of local observables after perturbations are thermal~\cite{dalessio_kafri_16} and the system satisfies eigenstate thermalization hypothesis (ETH)~\cite{deutsch_91, srednicki_94, rigol_dunjko_08, dalessio_kafri_16, eisert_friesdorf_15, deutsch_18, mori_ikeda_18}.
Exceptions from that generic behavior are currently under active investigation.
A well-established nonergodic behavior in translationally invariant systems occurs at integrable points~\cite{kinoshita_wenger_06, tang_kao_18, caux_essler_13, wouters_denardis_14, pozsgay_mestyan_14, ilievski15, Calabrese_2016, essler_fagotti_2016, cazalilla_chung_2016}, for which eigenstates do not obey ETH~\cite{pop1, Cassidy2011, steinigeweg2013, Ikeda2013, Alba2015, rigol_16, vidmar16, leblond_mallayya_19, mierzejewski_vidmar_20} and the long-time averages of observables after quantum quenches are described by the generalized Gibbs ensemble~\cite{gge, vidmar16}.
Recently, nonergodic properties of certain eigenstates (termed many-body scars) were observed in some translationally invariant models away from integrable points~\cite{Shiraishi2017, mondaini_mallayya_18, moudgalya_regnault_18, turner_michailidis_18, turner_michailidis_18_a, lin_motrunich_19, iadecola_znidaric_19, schecter_iadecola_19}.
For disordered quantum many-body systems in one dimension that are the focus of this Letter, it is proposed that (almost) {\it all} eigenstates become nonergodic at large enough disorder due to localization~\cite{oganesyan_huse_07, pal_huse_10}.
This scenario predicts an eigenstate quantum phase transition from an ergodic to a nonergodic phase~\cite{pal_huse_10, bauer_nayak_13, huse_nandkishore_13, pekker_refael_14}, the latter named many-body localization (MBL)~\cite{basko_aleiner_06, gornyi_mirlin_05, Rahul15, altman_vosk_15, abanin_altman_19}.

MBL and the corresponding transition were widely studied by means of numerical approaches~\cite{oganesyan_huse_07, kjall_bardarson_14, devakul_singh_15, Luitz2015, bertrand_garciagarcia_16, serbyn_moore_16, luitz_16, yu_luitz_16, khemani_lim_17, khemani_sheng_17, lin_sbierski_18, buijsman_cheianov_19, sierant_zakrzewski_19, sierant_zakrzewski_20, barkelbach_reichman_10, barisic_prelovsek_10, khatami_rigol_12, modak_mukerjee_15, bera_schomerus_15, mondaini_rigol_15, torresherrera_santos_15, serbyn_papic_15, serbyn_papic_17, barisic_kokalj_16, detomasi_bera_17, doggen_schindler_18, mace_alet_19, lenarcic_alberton_20, guidici_surace_20, corps_molina_20} and phenomenological theories~\cite{Serbyn2013, huse14, ros15, chandran_pal_16, pekker_clark_17, vosk_huse_15, potter_vasseur_15, zhang_zhao_16, dumitrescu_vasseur_17, deroeck_huveneers_17, thiery_huveneers_18, balasubramanian_liao_20}, as well as experimentally~\cite{schreiber_hodgman_15, kondov_mcgehee_15, smith_lee_16, choi_hild_16, lukin_rispoli_19, rispoli_lukin_19, chiaro_neill_20}.
Moreover, under certain assumptions in rigorous approaches, there are arguments about existence of the MBL phase~\cite{imbrie_16, imbrie_16_short}.
The nature of the transition, however, remains less clear.
Numerically, the transition was mainly studied within the framework of power-law divergence of correlation length~\cite{kjall_bardarson_14, Luitz2015, bertrand_garciagarcia_16, khemani_sheng_17},
\begin{equation} \label{def_xi_powerlaw}
 \xi_0 = \frac{1}{|W-W^*|^\nu} \,,
\end{equation}
where $W^*$ is the critical disorder and $\nu$ is the critical exponent.
The main concern with most of the numerical results in systems with uncorrelated disorder is that they suggest $\nu \sim 1$, which violates the Harris bound $\nu \geq 2$~\cite{harris_74, chayes_chayes_86, chandran_laumann_15}.
In contrast, phenomenological approaches based on real-space renormalization group (RG) typically predict $\nu > 2$~\cite{vosk_huse_15, potter_vasseur_15, zhang_zhao_16, dumitrescu_vasseur_17}.
Recently, modified RG schemes~\cite{dumitrescu_goremykina_19, goremykina_vasseur_19} proposed a Kosterlitz-Thouless (KT) type of transition with a correlation length that depends on the disorder as
\begin{equation} \label{def_xi_kt}
 \xi_{\rm KT} = \exp\left\{\frac{b_\pm}{\sqrt{|W-W^*|}}\right\} \, ,
\end{equation}
where $b_-$ ($b_+$) are nonuniversal parameters below (above) the transition.
Hence, there is currently a gap between predictions of exact numerical and phenomenological approaches and as a consequence, the nature of the transition remains an open problem.

A new perspective in understanding of the ergodicity breaking transition was recently obtained by calculating the spectral form factor~\cite{suntajs_bonca_20a}, 
whose finite-size dependence was interpreted as a linear drift of the transition point with system size, $W^* \propto L$, for system sizes where numerical diagonalization of the full Hamiltonian matrix is accessible.
This result raised the question whether such linear drift is an asymptotic feature suggesting that the transition to MBL is a crossover, or a preasymptotic feature consistent with a phase transition at some very large value of disorder in the thermodynamic limit.
Subsequent work~\cite{abanin_bardarson_20, sierant_delande_20, panda_scardicchio_20} mostly argued in favor of the second option.
It is therefore an urgent need to introduce new unbiased numerical measures to characterize the transition, which should also provide a benchmark for subsequent phenomenological studies.

The goal of this Letter is to quantitatively compare different scenarios of the ergodicity breaking transition.
We introduce a cost function approach to describe the quality of the finite-size data collapse of ergodicity indicators as functions of $L/\xi$ (i.e., the system size $L$ divided by the correlation length $\xi$).
This approach enables us to extract the most optimal form of the correlation length and to locate the disorder transition point in finite systems. 
For the numerically accessible system sizes, our results consistently exhibit two main features:
the correlation length $\xi$ follows the Kosterlitz-Thouless behavior~(\ref{def_xi_kt}), and ergodicity indicators at the transition are very close to random-matrix theory predictions.

{\it Model and Methodology.}
We study interacting spin-1/2 Hamiltonians with on-site disorder on a one-dimensional periodic lattice with $L$ sites,
\begin{align} \label{def_Ham}
 \hat H & = \sum_{j=1}^2  J_j \sum_{\ell=1}^{L} \left( \hat s_{\ell}^x \hat s_{\ell+j}^x + \hat s_{\ell}^y \hat s_{\ell+j}^y + \Delta_j \hat s_{\ell}^z \hat s_{\ell+j}^z \right)
 + \sum_{\ell=1}^L w_{\ell} \hat s_\ell^z \, ,
\end{align}
where $\hat s_\ell^\alpha$ ($\alpha = x,y,z$) are spin-1/2 operators at site $\ell$.
We consider the total spin projection $s^z=0$ sector and set $J_1 \equiv 1$ as the unit of energy.
Disorder with the magnitude $W$ is generated by independent and identically distributed local magnetic fields, with values $w_\ell \in [-W,W]$ drawn from a uniform distribution.

We study two disordered models, the $J_1$-$J_2$ model ($\Delta_1 = \Delta_2 = 0.55$, $J_2=1$), and the Heisenberg model ($\Delta_1=1$ and $J_2 = 0$).
For a given disorder distribution $\{ w_\ell \}$, we calculate exact eigenstates around the center of the spectrum using shift and invert diagonalization method~\cite{pietracaprina2018shift}.

We focus on two widely studied ergodicity indicators that characterize properties of Hamiltonian eigenstates and eigenvalues: the eigenstate entanglement entropy $S$ and the spectral level spacing ratio $r$, respectively.
In the context of Anderson localization, statistics of Hamiltonian eigenvalues and the corresponding scaling solutions as functions of $L/\xi_0$ represented one of the main numerical approaches to detect the transition point~\cite{shklovskii_shapiro_93, zharekeshev_kramer_97}.
Recently, ergodicity indicators $S$ and $r$ have been extensively studied in the context of disordered interacting spin chains~\cite{oganesyan_huse_07, bauer_nayak_13, grover_14, kjall_bardarson_14, devakul_singh_15, Luitz2015, bertrand_garciagarcia_16, serbyn_moore_16, luitz_16, yu_luitz_16, khemani_lim_17, khemani_sheng_17, lin_sbierski_18, buijsman_cheianov_19, sierant_zakrzewski_19, sierant_zakrzewski_20, richter_schubert_20, barlev_cohen_15}.

\begin{figure}[!]
\includegraphics[width=1.00\columnwidth]{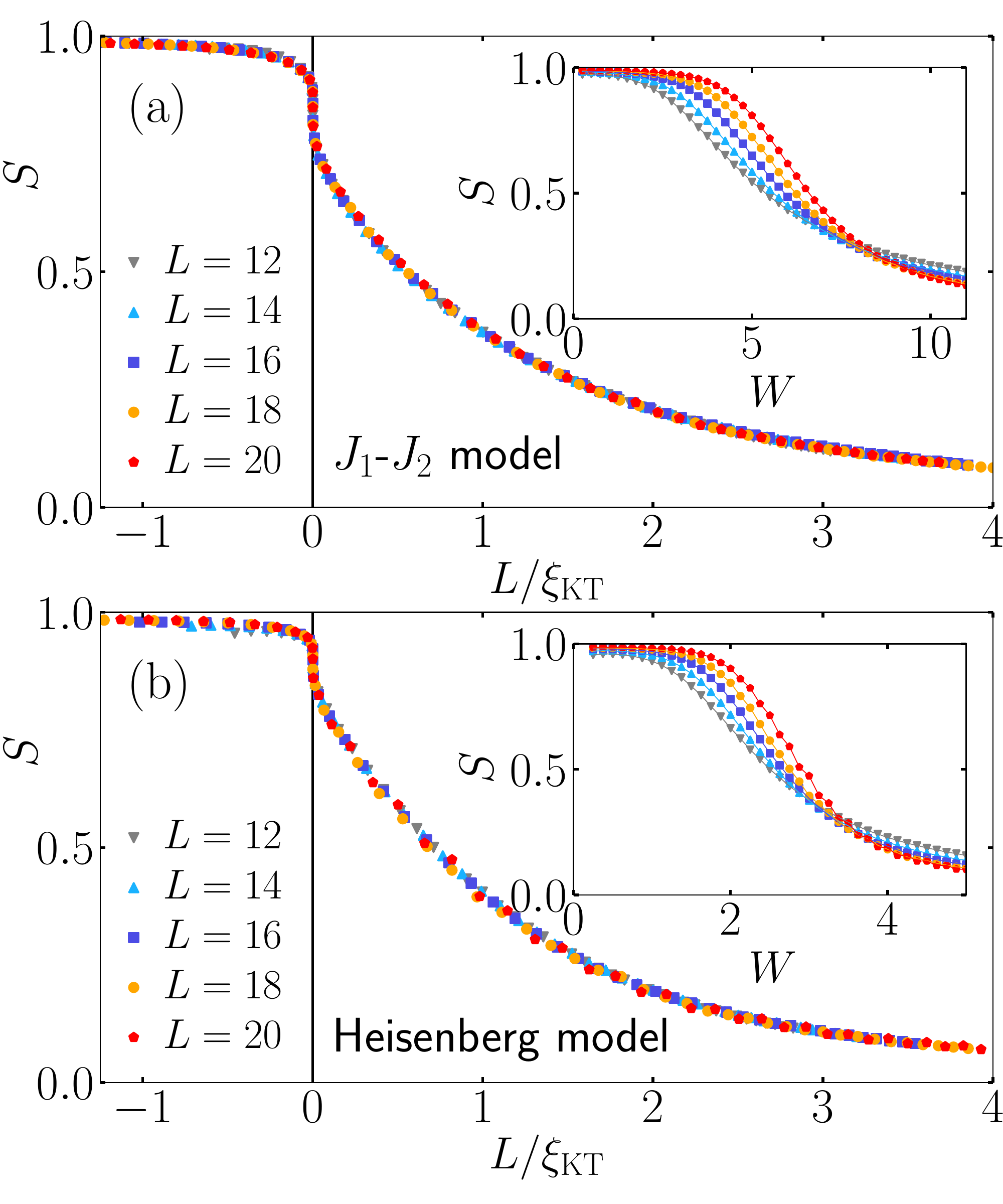}
\caption{
Eigenstate entanglement entropy $S$ for different systems sizes $L$, calculated (a) in the $J_1$-$J_2$ model and (b) in the Heisenberg model.
Insets show $S$ as a function of disorder $W$.
In the main panels, we plot $S$ as a function of $L/\xi_{\rm KT}$ [$-L/\xi_{\rm KT}$ if $W<W^*$], where $\xi_{\rm KT}$ is a KT correlation length~(\ref{def_xi_kt}), assuming $b_- = b_+ \equiv b$ and the transition point ansatz $W^* = w_0 + w_1 L$.
The optimal parameters $b$, $w_0$ and $w_1$ in $\xi_{\rm KT}$ are obtained by minimizing the cost function ${\cal C}_S(\xi_{\rm KT})$ in Eq.~(\ref{def_costfun}).
The number of data points included in the minimization procedure is $N_{\rm p}=287$ in panel (a) and $N_{\rm p}=225$ in panel (b).
See Fig.~\ref{fig_Wcritical} and~\cite{suppmat} for details.
}
\label{fig_SvN}
\end{figure}

\begin{figure}[!]
\includegraphics[width=1.00\columnwidth]{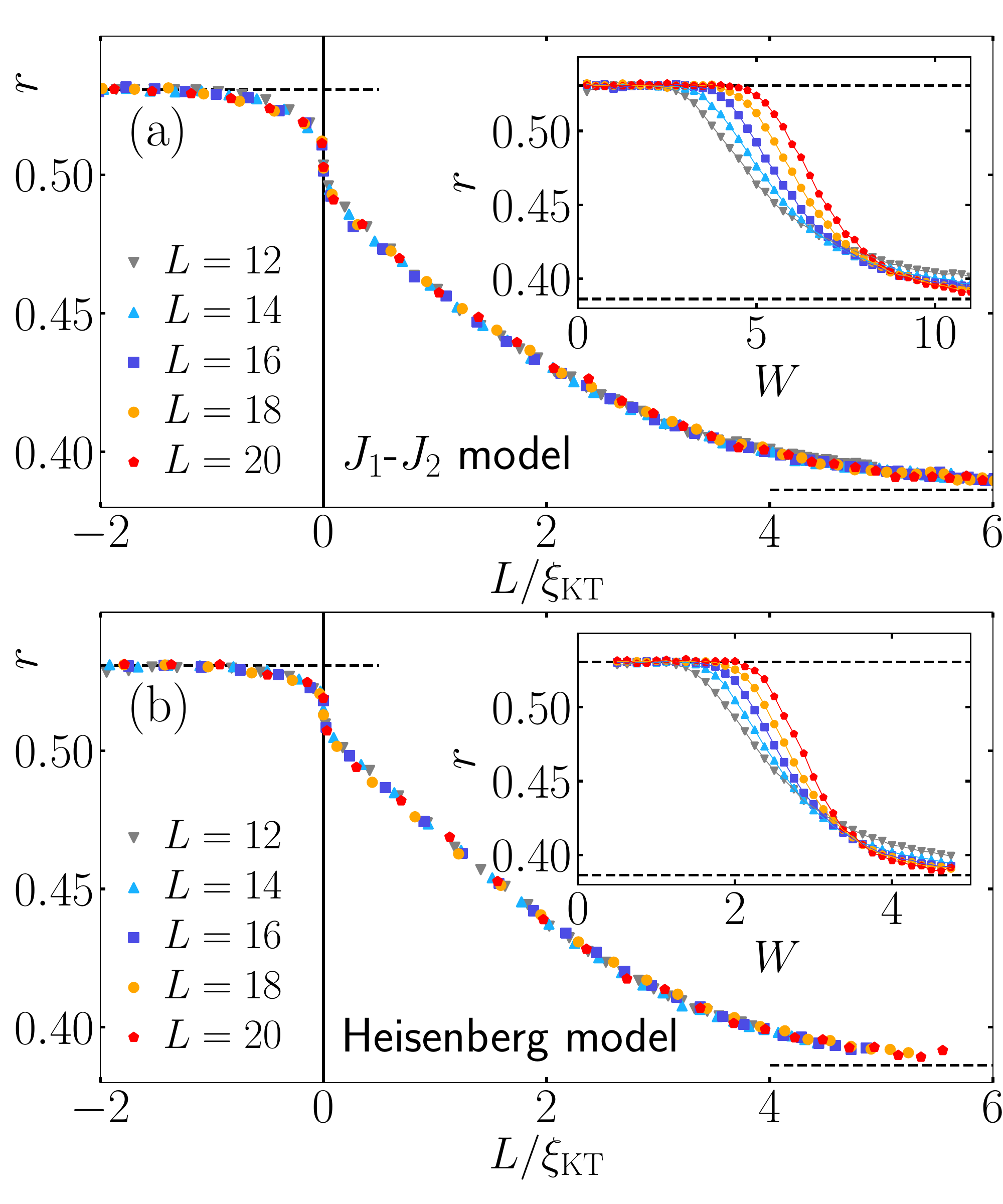}
\caption{
Level spacing ratio $r$ for different systems sizes $L$, calculated (a) in the $J_1$-$J_2$ model and (b) in the Heisenberg model.
Insets show $S$ as a function of disorder $W$.
In the main panels, we plot $r$ as a function of $L/\xi_{\rm KT}$ [$-L/\xi_{\rm KT}$ if $W<W^*$], where $\xi_{\rm KT}$ is a KT correlation length~(\ref{def_xi_kt}), assuming $b_- = b_+ \equiv b$ and the transition point ansatz $W^* = w_0 + w_1 L$.
The optimal parameters $b$, $w_0$ and $w_1$ in $\xi_{\rm KT}$ are obtained by minimizing the cost function ${\cal C}_S(\xi_{\rm KT})$ in Eq.~(\ref{def_costfun}).
The number of data points included in the minimization procedure is $N_{\rm p}=285$ in panel (a) and $N_{\rm p}=175$ in panel (b).
See Fig.~\ref{fig_Wcritical} and~\cite{suppmat} for details.
}
\label{fig_r}
\end{figure}

First, we calculate the von Neumann entanglement entropy $S_\alpha =  -{\rm Tr}\{ \hat \rho_A \ln(\hat \rho_A) \}$ in an eigenstate $|\alpha\rangle$, where the subsystem $A$ consists of the first $L/2$ lattice sites, $\hat \rho_A = {\rm Tr}_{L-A} \{\hat \rho\}$ is the trace over the remaining $L/2$ sites, and $\hat \rho = |\alpha\rangle \langle \alpha|$.
Since we study the total $s^z=0$ sector, we divide $S_\alpha$ by the corresponding random-matrix theory (RMT) result $S_{\rm RMT} = (L/2) \ln(2) + (1/2 + \ln(1/2))/2  - 1/2$, which includes ${\cal O}(1)$ contributions and hence minimizes finite-size effects~\cite{vidmar_rigol_17}.
To calculate the level spacing ratio $r$, we first calculate $r_\alpha = \min\{\delta_\alpha, \delta_{\alpha-1}\}/\max\{\delta_\alpha, \delta_{\alpha-1}\}$ for an eigenstate $|\alpha\rangle$, where $\delta_\alpha = E_{\alpha+1} - E_{\alpha}$ is the energy level spacing~\cite{oganesyan_huse_07}.
We then obtain $S$ and $r$ by averaging $S_\alpha/S_{\rm RMT}$ and $r_\alpha$, respectively, over eigenstates around the center of the spectrum and over different realizations of the disorder distribution $\{ w_\ell\}$~\cite{suppmat}.
Results for $S(W)$ and $r(W)$ for different system sizes $L$ are shown for both models in the insets of Fig.~\ref{fig_SvN} and~\ref{fig_r}, respectively.
In the ergodic (small $W$) regime, $S\approx 1$ and $r \approx r_{\rm GOE} \approx 0.5307$~\cite{atas_bogomolny_13}, while at large $W$ in finite systems, $S \to 0$~\cite{bauer_nayak_13} and $r \to r_{\rm Poisson} = 2\ln(2)-1 \approx 0.3863$~\cite{oganesyan_huse_07}.

Our central goal is to find the best data collapse of $S(W,L)$ and $r(W,L)$ as functions of $L/\xi$.
Specifically, we want to establish an unbiased, quantitative measure of the quality of the data collapse for different functional forms of the correlation length $\xi$ [such as $\xi_0$ and $\xi_{\rm KT}$ from Eqs.~(\ref{def_xi_powerlaw})-(\ref{def_xi_kt})], and the critical disorder $W^*$, which is included in the expression for $\xi$.
This is achieved by introducing the cost function for a quantity $X \in \{S, r \}$ that consists of $N_p$ values at different $W$ and $L$,
\begin{equation} \label{def_costfun}
 {\cal C}_X = \frac{ \sum_{j=1}^{N_{\rm p}-1} |X_{j+1} - X_{j}| }{\max\{X_j\} - \min\{X_j\}} - 1 \,.
\end{equation}
In Eq.~(\ref{def_costfun}), we sort all $N_{\rm p}$ values of $X_j$ according to nondecreasing values of ${\rm sign}[W$-$W^*] L/\xi$.
In the case of an ideal data collapse, this implies $\sum_j |X_{j+1} - X_{j}| = \max\{X_j\} - \min\{X_j\}$ and therefore ${\cal C}_X = 0$.
For the large data sets studied here, the cost function is always positive, ${\cal C}_X > 0$.
Our goal is to find, for given functional forms of the correlation length $\xi(W,L)$ and the critical disorder $W^*(L)$, the optimal values of fitting parameters that minimize ${\cal C}_X$.
We apply the cost function minimization algorithm to results for $S(W,L)$ and $r(W,L)$ at $L=12,14,16,18,20$ in the $J_1$-$J_2$ model and the Heisenberg model (see Supplemental Material~\cite{suppmat} for details).

{\it Results.}
We now describe our main results, valid for both ergodicity indicators $S$ and $r$, and for both investigated models.
For simplicity, we consider a single fitting parameter $b_- = b_+ \equiv b$ in $\xi_{\rm KT}$ in Eq.~(\ref{def_xi_kt}).
The scenario with $b_- \neq b_+$ and the optimal values of $b$ are discussed in~\cite{suppmat}.

(i) For the simplest functional forms of the transition point we consider two fitting functions $W^* = w_0$ and $W^* = w_0 + w_1 L$, with free parameters $w_0$ and $w_1$.
If we only consider an $L$-independent function $W^* = w_0$, the data collapse (quantified in terms of the cost function) is better as a function of $L/\xi_0$ than $L/\xi_{\rm KT}$
[left columns in Tables~\ref{tab:table1} and~\ref{tab:table2}].
However, the data collapse becomes much better (i.e., the cost function becomes much lower) for the KT transition if $W^*$ is allowed to increase linearly with $L$ [central columns in Tables~\ref{tab:table1} and~\ref{tab:table2}].
The latter statement holds true also in the special case of zero offset, $W^* = w_1 L$.

As an example, we show in Fig.~\ref{fig_SvN} the entanglement entropy $S$ for both models using $W^* = w_0 + w_1 L$, and plot results in the main panels as functions of $L/\xi_{\rm KT}$.
In Fig.~\ref{fig_r}, analogous results are shown for the level spacing ratio $r$.
In both figures, the scaling collapses appear to be excellent.

Results at $L/\xi_{\rm KT} \approx 0$ suggests two remarkable observations. 
The first is an emergent discontinuity of the rescaled data at the transition.
Scenarios for the transition with discontinuous jumps of $S$ at the critical point were previously discussed in Refs.~\cite{khemani_lim_17, dumitrescu_vasseur_17}.
The second observation is that $r(L/\xi_{\rm KT} \approx 0)$ and $S(L/\xi_{\rm KT} \approx 0)$ are very close to the RMT predictions.
Hence, their values are close to the thermal values.

\begin{figure}[!]
\includegraphics[width=1.00\columnwidth]{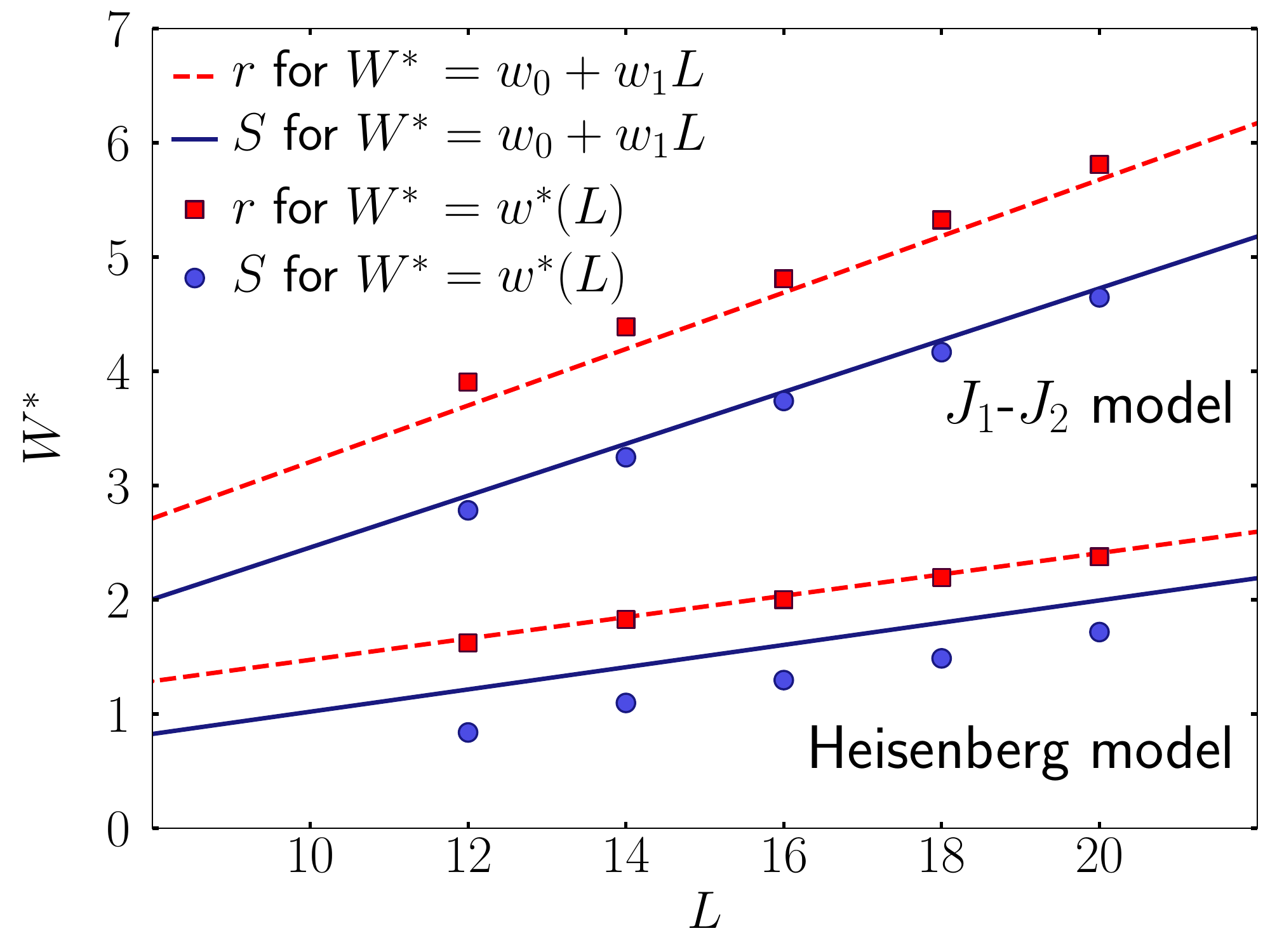}
\caption{
Ergodicity breaking transition point $W^*$ as a function of system size $L$, for the $J_1$-$J_2$ model (upper part) and the Heisenberg model (lower part).
Results for $W^*$ are obtained from the best data collapse using a KT correlation length $\xi_{\rm KT}$ from Eq.~(\ref{def_xi_kt}), with $b$ as a free parameter (assuming $b_- = b_+)$.
Lines are results for a transition point ansatz $W^* = w_0 + w_1 L$ (with free parameters $w_0$ and $w_1$),
symbols are results for a function-independent transition point $W^* = w^*(L)$ [with five free parameters for five different systems sizes $L=12,14,16,18,20$].
Results are shown for $S$ (circles and solid lines) and $r$ (squares and dashed lines).
In the $J_1$-$J_2$ model we get $(w_0, w_1) = (0.19,0.23)$ for $S$ and $(0.74,0.25)$ for $r$, while in the Heisenberg model we get $(w_0, w_1) = (0.05,0.10)$ for $S$ and $(0.54,0.09)$ for $r$.
Therefore, both ergodicity indicators $S$ and $r$ have very similar leading term $w_1$ within the same model, while the subleading term $w_0$ may be different.
}
\label{fig_Wcritical}
\end{figure} 

(ii) The results from (i) are robust towards more general functional forms of the transition point $W^*$.
First, we tested the fitting functions $W^* = w_0 + w_1/L$ and $W^* = w_0 + w_1/\ln(L)$, which are two parameter fits to $W^*$, but always yielded substantially higher cost function when compared to the function $W^* = w_0 + w_1 L$ (see~\cite{suppmat} for details).
Then, we tested a general ansatz $W^* = w^*(L)$ using independent fitting values of $w^*$ for every $L$ [i.e., five different $w^*$ for five different systems sizes $L=12,14,16,18,20$].
We find that for system sizes under investigation, the resulting values of $w^*(L)$ after minimization are rather accurately described by the function $W^* = w_0 + w_1 L$ for both ergodicity indicators $S$ and $r$, as shown in Fig.~\ref{fig_Wcritical}.

We compare the values of the cost functions for the disordered $J_1$-$J_2$ model and the disordered Heisenberg model in Tables~\ref{tab:table1} and~\ref{tab:table2}, respectively.
In Figs.~\ref{fig_Supp_S_J1J2}-\ref{fig_Supp_r_XXZ} of~\cite{suppmat} we also visually compare the scaling collapses of $S$ and $r$ with and without the drift in the functional dependence of the transition point $W^*$.
Note that the cost functions ${\cal C}_X[\xi_{\rm KT}]$ for the general model $W^*=w^*(L)$ take only slighly lower values than for the linear one $W^*=w_0+w_1 L$.

We interpret our results as evidence that a transition with a KT correlation length $\xi_{\rm KT}$ is more favorable than a transition with a power-law correlation length $\xi_0$.
A particularly suggestive evidence supporting the latter statement is that cost functions using $\xi_{\rm KT}$ with a two-parameter transition point function $W^* = w_0 + w_1 L$ [central columns in Tables~\ref{tab:table1} and~\ref{tab:table2}] are substantially lower than cost functions using $\xi_0$ with a five-parameter function $W^* = w^*(L)$ [right columns in Tables~\ref{tab:table1} and~\ref{tab:table2}].

Moreover, for the numerically available system sizes the best scaling collapse of the transition using a KT correlation length $\xi_{\rm KT}$ consistently exhibits a linear drift with system size, $W^* \propto L$, as suggested by Fig.~\ref{fig_Wcritical}.
However, an analogous statement cannot be made for the transition using $\xi_0$.
In the latter case, using $W^*=w^*(L)$ [right column in Tables~\ref{tab:table1} and~\ref{tab:table2}], we find degenerate solutions (i.e., almost identical values of cost functions) with very different functional forms of $W^*$.

The observed linear scaling $W^* \propto w_1 L$ opens a question about its fate in the thermodynamic limit.
While the symbols in Fig.~\ref{fig_Wcritical} show no tendency towards approaching a horizontal line, we can also not exclude scenarios where $W^*(L)\propto w_1 L$ represents a small size behavior that eventually saturates to a finite critical point $W_\infty^*$ in the thermodynamic limit.
As a quantitative estimate of the latter scenario we consider the functional form $W^*(L) = W_\infty^* \tanh{(L/L_0)}$, where $L_0$ represents a characteristic length scale.
This form reproduces the linear $L$ dependence at $L \ll L_0$ since $W^*(L) \approx W_\infty^* L/L_0 + {\cal O} (L^3/L_0^3)$, where $w_1 = W_\infty^*/L_0$.
Then, by requiring the leading term to be much larger than the subleading term at $L=20$, one could estimate a lower bound for $W_\infty^*$.
Using the cost function minimization approach, we estimate $L_0 \gtrsim 50$ (see~\cite{suppmat} for details), and hence $W_\infty^* \gtrsim 5$ in the Heisenberg model (using $w_1=0.10$) and $W_\infty^* \gtrsim 12$ in the $J_1$-$J_2$ model (using $w_1=0.24$).

{\it Discussion.}
Intriguingly, our analysis predicts nearly thermal (RMT-like) properties of the transition point in finite systems.
This is a consequence of the transition taking place at relatively weak disorder:
for $L\approx 20$, it occurs at $W^*\approx 2$ in the Heisenberg model and at $W^*\approx 5$ in the $J_1$-$J_2$ model.
These values are lower than those usually considered in the MBL literature, which mostly followed the initial proposal~\cite{oganesyan_huse_07} for the transition point having insulator-like (Poisson) statistics.
Indeed, our scenario for the transition differs from previous numerical studies of the same ergodicity indicators in the Heisenberg model~\cite{Luitz2015, bertrand_garciagarcia_16}, which explored the optimal data collapse as a function of $L/\xi_0$ and obtained $W^* \approx 3.4 - 3.7$ for comparable system sizes.

The outcome of our scaling analysis can be considered as a first step in unifying exact numerical calculations with RG approaches that also predict a KT-like transition~\cite{dumitrescu_goremykina_19, goremykina_vasseur_19}.
One of the next goals is to better understand the character of the transition point:
while our results suggest nearly thermal properties, the RG schemes predict a vanishing density of thermal blocks at the transition~\cite{dumitrescu_goremykina_19, goremykina_vasseur_19}.

\begin{table}[]
\caption{\label{tab:table1}
Cost function ${\cal C}_X$, see Eq.~(\ref{def_costfun}), in the $J_1$-$J_2$ model.
Values of ${\cal C}_X$ are shown for two ergodicity indicators $X \in \{ S, r\}$, using correlation lengths $\xi_0$ and $\xi_{\rm KT}$ from Eqs.~(\ref{def_xi_powerlaw})-(\ref{def_xi_kt}).
Columns denote different functional forms of $W^*$ used in $\xi_0$ and $\xi_{\rm KT}$.
Results are obtained using the data points in Figs.~\ref{fig_SvN}(a) and~\ref{fig_r}(a) for $W>0.5$.
}
\begin{ruledtabular}
\begin{tabular}{ l | c c c }
&
$W^* = w_0$ & $W^* = w_0 + w_1 L$ & $W^* = w^*(L)$\\
\colrule
${\cal C}_S[\xi_{\rm KT}]$ & 3.99 & 0.34 & 0.29 \\
${\cal C}_S[\xi_0]$ & 2.80 & 1.81 & 1.71 \\
\colrule
${\cal C}_r[\xi_{\rm KT}]$ & 5.46 & 1.01 & 0.92 \\
${\cal C}_r[\xi_0]$ & 4.31 & 2.71 & 2.57
\end{tabular}
\end{ruledtabular}
\end{table}

\begin{table}[]
\caption{\label{tab:table2}
Cost function ${\cal C}_X$, see Eq.~(\ref{def_costfun}), in the Heisenberg model.
Values of ${\cal C}_X$ are shown for two ergodicity indicators $X \in \{ S, r\}$, using correlation lengths $\xi_0$ and $\xi_{\rm KT}$ from Eqs.~(\ref{def_xi_powerlaw})-(\ref{def_xi_kt}).
Columns denote different functional forms of $W^*$ used in $\xi_0$ and $\xi_{\rm KT}$.
Results are obtained using the data points in Figs.~\ref{fig_SvN}(b) and~\ref{fig_r}(b) for $W>0.5$.
}
\begin{ruledtabular}
\begin{tabular}{ l | c c c }
&
$W^* = w_0$ & $W^* = w_0 + w_1 L$ & $W^* = w^*(L)$\\
\colrule
${\cal C}_S[\xi_{\rm KT}]$ & 2.51 & 0.46 & 0.29 \\
${\cal C}_S[\xi_0]$ & 1.80 & 0.94 & 0.77 \\
\colrule
${\cal C}_r[\xi_{\rm KT}]$ & 2.84 & 0.51 & 0.46 \\
${\cal C}_r[\xi_0]$ & 2.16 & 1.08 & 1.01
\end{tabular}
\end{ruledtabular}
\end{table}

Our analysis applies to disorder averages of ergodicity indicators.
When the disorder is increased, fluctuations of ergodicity indicators (at least in finite systems) may become anomalous, which has been observed in fluctuations of the entanglement entropy~\cite{kjall_bardarson_14, luitz_16, yu_luitz_16, khemani_lim_17, khemani_sheng_17} and in distributions of other observables~\cite{serbyn_papic_17, sahu_xu_19, colmenarez_mcclarty_19, schulz_taylor_20, mierzejewski_sroda_20}.
It remains open how the KT character of the transition of averaged quantities is related to other statistical properties of the model.

{\it Conclusions.}
In this Letter we proposed an unbiased, quantitative approach based on cost function minimization to test the nature of a disorder driven transition in finite quantum spin chains.
We argued that certain key ergodicity indicators exhibit clear signatures of the ergodicity breakdown and remarkable finite-size scaling properties, which do not violate the Harris bound~\cite{harris_74, chayes_chayes_86, chandran_laumann_15} and are consistent with the KT quantum phase transition.

Moreover, the cost function minimization approach for numerically accessible system sizes results in two features: a nearly thermal character of the transition point and its linear drift with the system size. 
In fact, both properties emerge simultaneously with the KT character of the transition.
The main open question is how close are the numerical results to the true asymptotic regime, and what is the relation between the observed properties of the transition point and the KT nature of the transition when the thermodynamic limit is approached.
More work is needed to clarify this.

\acknowledgements
We acknowledge discussions with F. Heidrich-Meisner.
This work is supported by the Slovenian Research Agency (ARRS), Research core fundings No.~P1-0044 (J.\v S., J.B. and L.V.), No.~P1-0402 (T.P.) and No.~J1-1696 (L.V.), and by the European Research Council (ERC) under Advanced Grant 694544 -- OMNES (T.P.).

\bibliographystyle{biblev1}
\bibliography{references}


\newpage
\phantom{a}
\newpage
\setcounter{figure}{0}
\setcounter{equation}{0}
\setcounter{table}{0}

\renewcommand{\thetable}{S\arabic{table}}
\renewcommand{\thefigure}{S\arabic{figure}}
\renewcommand{\theequation}{S\arabic{equation}}

\renewcommand{\thesection}{S\arabic{section}}

\onecolumngrid

\begin{center}

{\large \bf Supplemental Material:\\
Ergodicity Breaking Transition in Finite Disordered Spin Chains}\\

\author{Jan \v Suntajs}
\affiliation{Department of Theoretical Physics, J. Stefan Institute, SI-1000 Ljubljana, Slovenia}
\author{Janez Bon\v ca}
\affiliation{Department of Physics, Faculty of Mathematics and Physics, University of Ljubljana, SI-1000 Ljubljana, Slovenia}
\affiliation{Department of Theoretical Physics, J. Stefan Institute, SI-1000 Ljubljana, Slovenia}
\author{Toma\v z Prosen}
\affiliation{Department of Physics, Faculty of Mathematics and Physics, University of Ljubljana, SI-1000 Ljubljana, Slovenia}
\author{Lev Vidmar}
\affiliation{Department of Theoretical Physics, J. Stefan Institute, SI-1000 Ljubljana, Slovenia}
\affiliation{Department of Physics, Faculty of Mathematics and Physics, University of Ljubljana, SI-1000 Ljubljana, Slovenia}

\vspace{0.3cm}

Jan \v Suntajs$^{1}$, Janez Bon\v ca$^{2,1}$, Toma\v z Prosen$^{2}$ and Lev Vidmar$^{1,2}$\\
$^1${\it Department of Theoretical Physics, J. Stefan Institute, SI-1000 Ljubljana, Slovenia} \\
$^2${\it Department of Physics, Faculty of Mathematics and Physics, University of Ljubljana, SI-1000 Ljubljana, Slovenia}

\end{center}

\vspace{0.6cm}

\twocolumngrid

\label{pagesupp}

\section{Details about the cost function minimization procedure} \label{app1}

Here we provide more details about the minimization procedure of the cost function ${\cal C}_X$ introduced in Eq.~(\ref{def_costfun}) of the main text.

We first implement numerical exact diagonalization to calculate eigenstates and eigenvalues of the Hamiltonian under investigation.
For each Hamiltonian realization $\hat H_\mu$  (i.e., for fixed model parameters and for randomly generated disorder potentials), we target $N_{\rm eig}$ eigenstates closest to the mean energy $\bar E_\mu = {\rm Tr}\{\hat H_\mu\}/{\cal D}$, where ${\cal D} =$ is the Hilbert space dimension in the $s^z = 0$ sector.
We set $N_{\rm eig} = 100$ in Fig.~\ref{fig_SvN} and $N_{\rm eig} = 500$ in Fig.~\ref{fig_r}.
Results are further averaged over $N_{\rm sample}$ different realizations of disorder.
For the entanglement entropy $S$ in Fig.~\ref{fig_SvN}(a), we use $N_{\rm sample} \geq 1000$ for $L \leq 18$ and $N_{\rm sample} \geq 400$ for $L=20$,
while in Fig.~\ref{fig_SvN}(b), we use $N_{\rm sample} \geq 1000$ for $L \leq 18$ and $N_{\rm sample} \geq 100$ for $L=20$.
For the level spacing ratio $r$ in Fig.~\ref{fig_r}(a), we use $N_{\rm sample} \geq 450$ for $L \leq 18$ and $N_{\rm sample} \geq 350$ for $L=20$,
while in Fig.~\ref{fig_r}(b), we use $N_{\rm sample} \geq 1000$ for $L \leq 18$ and $N_{\rm sample} \geq 100$ for $L=20$.

For a given functional form of the correlation length $\xi$ and the critical disorder $W^*$ [which is included in the functional form of $\xi$, see Eqs.~(\ref{def_xi_powerlaw}) and~(\ref{def_xi_kt})], we then sort numerical values of $S$ and $r$ at different $W$ and $L \in \{12,14,16,18,20\}$ according to nondecreasing $L/\xi$.
In Fig.~\ref{fig_SvN}(a) the data included in the minimization procedure were for $0.5 \leq W \leq 15$ (except for for $L=20$, where $0.5\leq W \leq 13$) and $\Delta W = 0.25$.
In Fig.~\ref{fig_SvN}(b) the data included were for $0.5 \leq W \leq 6$ and $\Delta = 0.125$.
In Fig.~\ref{fig_r}(a) the data included in the minimization procedure were for $0.5 \leq W \leq 15$ (except for $L=20$, where $0.5 \leq W \leq 12.5$) and $\Delta W = 0.25$.
In Fig.~\ref{fig_r}(b) the data included were for $0.5 \leq W \leq 4.75$ and $\Delta = 0.125$.

We apply a differential evolution method implemented in {\tt scipy} to find the optimal set of free parameters of $\xi$ and $W^*$ that minimize the cost function.
In each realization, we employ a population size $10^2$ and allow for up to $10^3$ iterations.
We use the relative tolerance of convergence $10^{-2}$ and employ $10^2$ realizations of the algorithm to verify the precision of fitted parameters of the optimal solution.

In the main text (see Tables~\ref{tab:table1} and~\ref{tab:table2}) we compared cost functions with the functional form of the critical disorder $W^*=w_0$ [one free parameter], $W^* = w_0 + w_1 L$ [two free parameters] and $W^* = w^*(L)$ [five free parameters for five different $L$].
Here we complement those results by focusing on two-parameter functional forms $W^* = w_0 + w_1/L$ and $W^* = w_0 + w_1/\ln(L)$.
These two functions imply finite critical disorder $W^* = w_0$ in the thermodynamic limit $L \to \infty$.
Results for the optimal cost function are compared to the results for the functional form with a linear drift with $L$, i.e., $W^* = w_0 + w_1 L$, see Tables~\ref{tab:table1X} and~\ref{tab:table2X}.
The main result is that the functional form of the critical disorder $W^*$ always yields a lower cost function if a linear drift with $L$ is allowed.
Moreover, in the case of a linear drift the solution using the KT correlation length $\xi_{\rm KT}$ is always substantially better.
In other cases, there is no considerable difference in cost functions between solutions using $\xi_{\rm KT}$ or $\xi_0$.

\begin{table}[]
\caption{\label{tab:table1X}
Cost function ${\cal C}_X$, see Eq.~(\ref{def_costfun}), in the $J_1$-$J_2$ model.
Values of ${\cal C}_X$ are shown for two ergodicity indicators $X \in \{ S, r\}$, using correlation lengths $\xi_0$ and $\xi_{\rm KT}$ from Eqs.~(\ref{def_xi_powerlaw})-(\ref{def_xi_kt}).
Columns denote different functional forms of $W^*$ used in $\xi_0$ and $\xi_{\rm KT}$.
Results are obtained using the data points in Figs.~\ref{fig_SvN}(a) and~\ref{fig_r}(a) for $W>0.5$.
}
\begin{ruledtabular}
\begin{tabular}{ l | c c c }
\hspace*{0.3cm}$W^*$&
$ w_0 + \frac{w_1}{L}$ & $w_0 + \frac{w_1}{\ln(L)}$ & $ w_0 + w_1 L$\\
\colrule
${\cal C}_S[\xi_{\rm KT}]$ & 1.60 & 1.40 & 0.34 \\
${\cal C}_S[\xi_0]$ & 2.60 & 2.50 & 1.81 \\
\colrule
${\cal C}_r[\xi_{\rm KT}]$ & 2.73 & 2.38 & 1.01 \\
${\cal C}_r[\xi_0]$ & 3.72 & 3.27 & 2.71
\end{tabular}
\end{ruledtabular}
\end{table}

\begin{table}[]
\caption{\label{tab:table2X}
Cost function ${\cal C}_X$, see Eq.~(\ref{def_costfun}), in the Heisenberg model.
Values of ${\cal C}_X$ are shown for two ergodicity indicators $X \in \{ S, r\}$, using correlation lengths $\xi_0$ and $\xi_{\rm KT}$ from Eqs.~(\ref{def_xi_powerlaw})-(\ref{def_xi_kt}).
Columns denote different functional forms of $W^*$ used in $\xi_0$ and $\xi_{\rm KT}$.
Results are obtained using the data points in Figs.~\ref{fig_SvN}(b) and~\ref{fig_r}(b) for $W>0.5$.
}
\begin{ruledtabular}
\begin{tabular}{ l | c c c }
\hspace*{0.3cm}$W^*$&
$ w_0 + \frac{w_1}{L}$ & $w_0 + \frac{w_1}{\ln(L)}$ & $ w_0 + w_1 L$\\
\colrule
${\cal C}_S[\xi_{\rm KT}]$ & 1.31 & 1.08 & 0.46 \\
${\cal C}_S[\xi_0]$ & 1.33 & 1.36 & 0.94 \\
\colrule
${\cal C}_r[\xi_{\rm KT}]$ & 1.58 & 1.37 & 0.51 \\
${\cal C}_r[\xi_0]$ & 1.73 & 1.31 & 1.08
\end{tabular}
\end{ruledtabular}
\end{table}

\section{Qualitative comparison} \label{app2}

\begin{figure*}[!]
\includegraphics[width=2.00\columnwidth]{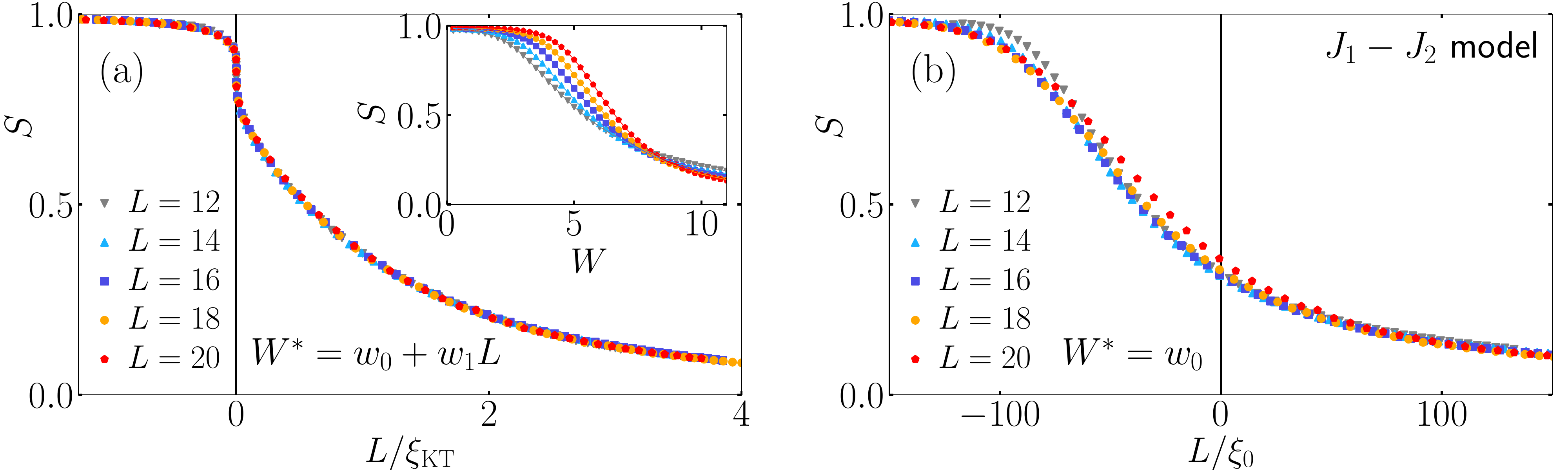}
\caption{
Eigenstate entanglement entropy $S$ for different systems sizes $L$ in the $J_1$-$J_2$ model.
We plot results as a function of $L/\xi$, using $\xi = \xi_{\rm KT}$ in (a) [assuming $b_- = b_+ \equiv b$ in Eq.~(\ref{def_xi_kt}), as in Fig.~\ref{fig_SvN}(a)], and $\xi = \xi_0$ in (b).
We use the transition point ansatz $W^* = w_0 + w_1 L$ in (a) and $W^* = w_0$ in (b).
The inset shows results as a function of disorder $W$.
Values of the cost function are ${\cal C}_S(\xi_{\rm KT})=0.34$ in (a) and ${\cal C}_S(\xi_0)=2.80$ in (b).
}
\label{fig_Supp_S_J1J2}
\end{figure*}

\begin{figure*}[!]
\includegraphics[width=2.00\columnwidth]{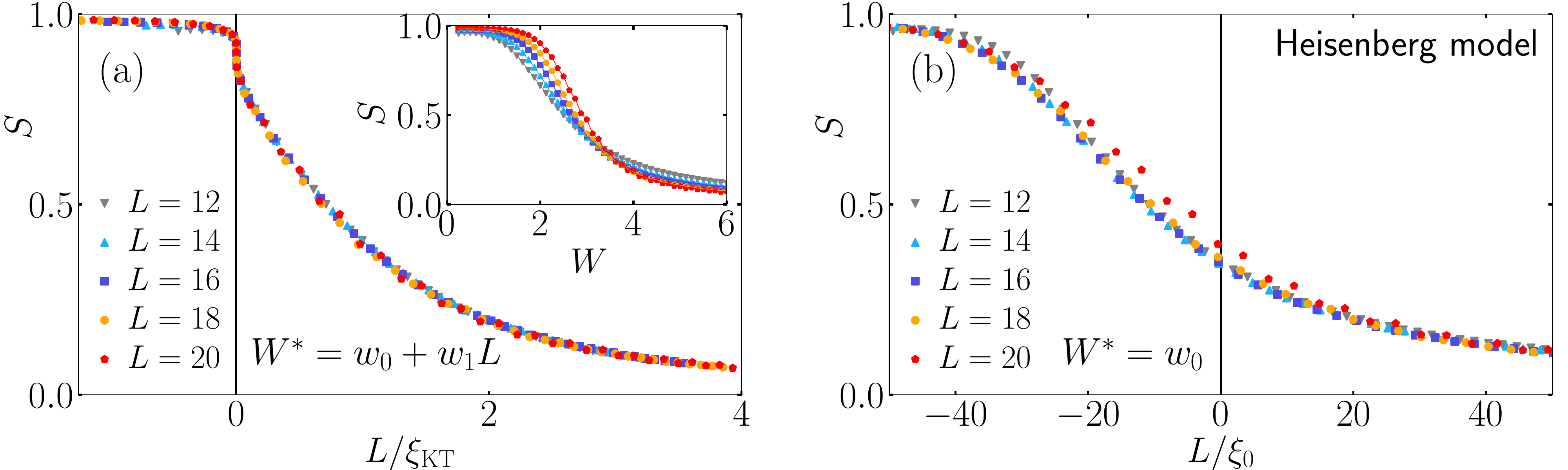}
\caption{
Eigenstate entanglement entropy $S$ for different systems sizes $L$ in the Heisenberg model.
We plot results as a function of $L/\xi$, using $\xi = \xi_{\rm KT}$ in (a) [assuming $b_- = b_+ \equiv b$ in Eq.~(\ref{def_xi_kt}), as in Fig.~\ref{fig_SvN}(b)], and $\xi = \xi_0$ in (b).
We use the transition point ansatz $W^* = w_0 + w_1 L$ in (a) and $W^* = w_0$ in (b).
The inset shows results as a function of disorder $W$.
Values of the cost function are ${\cal C}_S(\xi_{\rm KT})=0.46$ in (a) and ${\cal C}_S(\xi_0)=1.80$ in (b).
}
\label{fig_Supp_S_XXZ}
\end{figure*}

\begin{figure*}[!]
\includegraphics[width=2.00\columnwidth]{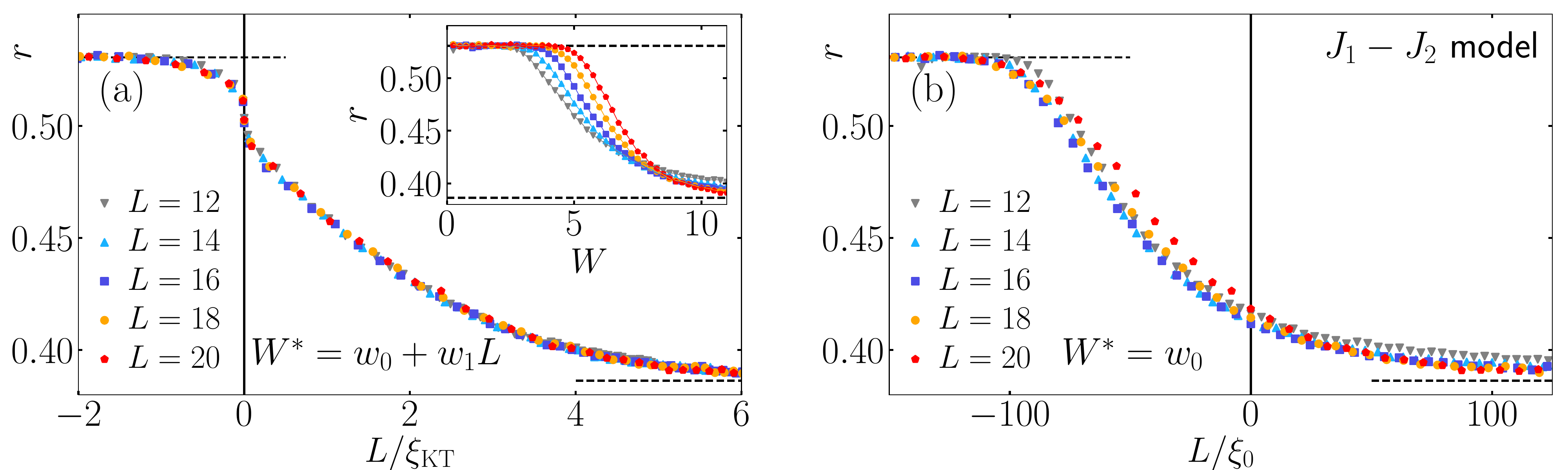}
\caption{
Level spacing ratio $r$ for different systems sizes $L$ in the $J_1$-$J_2$ model.
We plot results as a function of $L/\xi$, using $\xi = \xi_{\rm KT}$ in (a) [assuming $b_- = b_+ \equiv b$ in Eq.~(\ref{def_xi_kt}), as in Fig.~\ref{fig_r}(a)], and $\xi = \xi_0$ in (b).
We use the transition point ansatz $W^* = w_0 + w_1 L$ in (a) and $W^* = w_0$ in (b).
The inset shows results as a function of disorder $W$.
Values of the cost function are ${\cal C}_r(\xi_{\rm KT})=1.01$ in (a) and ${\cal C}_r(\xi_0)=4.31$ in (b).
}
\label{fig_Supp_r_J1J2}
\end{figure*}

\begin{figure*}[!]
\includegraphics[width=2.00\columnwidth]{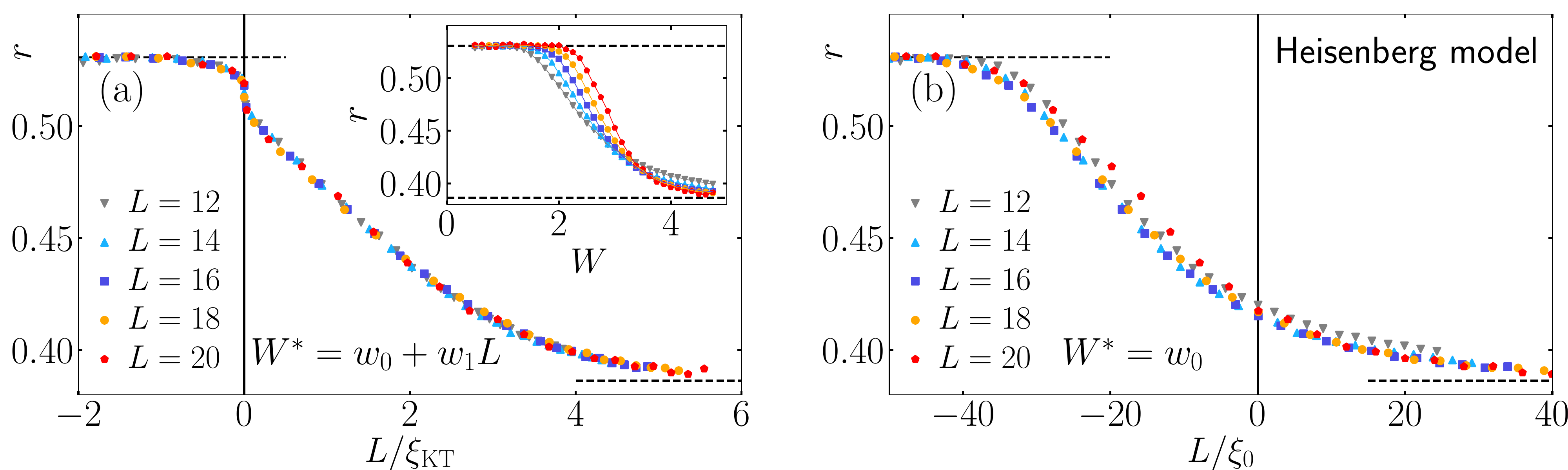}
\caption{
Level spacing ratio $r$ for different systems sizes $L$ in the Heisenberg model.
We plot results as a function of $L/\xi$, using $\xi = \xi_{\rm KT}$ in (a) [assuming $b_- = b_+ \equiv b$ in Eq.~(\ref{def_xi_kt}), as in Fig.~\ref{fig_r}(b)], and $\xi = \xi_0$ in (b).
We use the transition point ansatz $W^* = w_0 + w_1 L$ in (a) and $W^* = w_0$ in (b).
The inset shows results as a function of disorder $W$.
Values of the cost function are ${\cal C}_r(\xi_{\rm KT})=0.51$ in (a) and ${\cal C}_r(\xi_0)=2.16$ in (b).
}
\label{fig_Supp_r_XXZ}
\end{figure*}

Tables~\ref{tab:table1}-\ref{tab:table2} in the main text show quantitative comparison of the cost functions for different functional forms of the correlation length $\xi$ and critical disorder $W^*$.
Here we complement these results by showing qualitative (visual) comparison of the best data collapses.
We focus on functional forms of the critical disorder with the linear drift $W^* \propto L$ and in the absence thereof.
In the first case, we show the best data collapse as functions of $L/\xi_{\rm KT}$ using $W^*=w_0+w_1 L$, see Figs.~\ref{fig_Supp_S_J1J2}(a)-\ref{fig_Supp_r_XXZ}(a).
In the second case, we show the best data collapse as functions of $L/\xi_0$ using $W^*=w_0$, see Figs.~\ref{fig_Supp_S_J1J2}(b)-\ref{fig_Supp_r_XXZ}(b).
The results agree with expectations that the better data collapse in terms of the cost function also yields a visually more convincing data collapse.

\section{Lower bound estimate for the critical point} \label{app3}

In the main text we discussed scenarios for the large-$L$ dependence of the transition point $W^*(L)$.
We argued that our analysis may provide an estimate for the lower bound of $W^*$ if one assumes that the deviation from the linear drift $W^* \propto w_1 L$ emerges at system sizes that are only slightly larger that the maximal system size studied here, $L=20$.

\begin{figure*}[!]
\includegraphics[width=2.00\columnwidth]{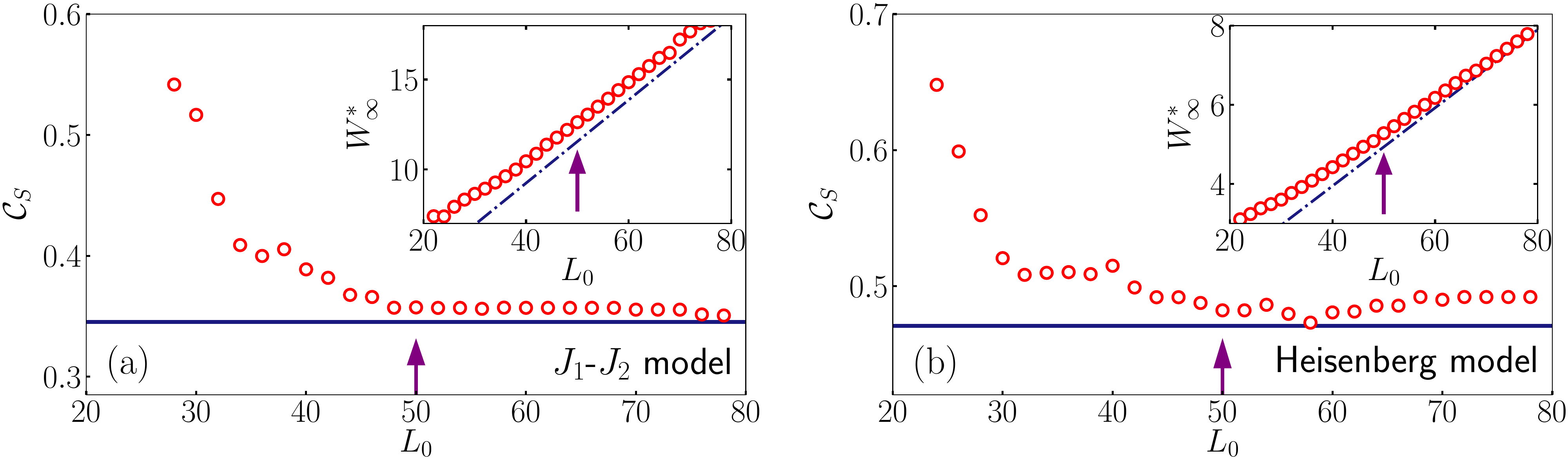}
\caption{
Cost functions ${\cal C}_S$ for $S$ in the $J_1$-$J_2$ model (a) and in the Heisenberg model (b), as a function of $L_0$.
We use the KT correlation length $\xi_{\rm KT}$ and the fitting function for the transition point $W^*(L) = W_\infty^* \tanh(L/L_0)$.
Horizontal lines represent the value $w_1$ if one uses the fitting function $W^*(L) = w_1 L$, and the arrows sketch the onset of roughly $L_0$-independent cost functions ${\cal C}_S$.
Insets: the optimal values of $W_\infty^*$ as a function of $L_0$.
Dashed lines represent $w_1 L_0$.
}
\label{fig_Wcritical_tanh}
\end{figure*}

For a quantitative analysis we use the fitting function for the transition point $W^*(L) = W_\infty^* \tanh(L/L_0)$.
We apply the cost function minimization algorithm for $S$ using $\xi_{\rm KT}$ with two free parameters $W_\infty^*$ and $b$ (i.e., $b_-=b_+$), while we fix $L_0$.
The corresponding cost function ${\cal C}_S$ as a function of $L_0$ is shown for both models in Fig.~\ref{fig_Wcritical_tanh}.
We define $\overline{L_0}$ as the lower bound for $L_0$ by requiring that ${\cal C}_S(L_0 \gtrsim \overline{L_0})$ is independent of $L_0$.
While the extraction of such a lower bound is less ambiguous for the $J_1$-$J_2$ model than for the Heisenberg model, we assume for both models $\overline{L_0} \approx 50$.
We use this value to estimate the lower bounds for the critical point listed in the main text.

\section{Different forms of the KT correlation length} \label{app4}

In the main text, we studied the scaling collapses of $S$ and $r$ using the KT correlation length $\xi_{\rm KT}$ from Eq.~(\ref{def_xi_kt}) with the identical parameter $b = b_- = b_+$ on both sides of the transition.
In the scaling analysis of $S$ in the main text, we get $b=4.87$ [$b=3.21$] in Fig.~\ref{fig_SvN}(a) [\ref{fig_SvN}(b)], and in the scaling analysis of $r$, we get $b=3.07$ [$b=1.96$] in Fig.~\ref{fig_r}(a) [\ref{fig_r}(b)].
Here we discuss the more general scenario when $b_- \neq b_+$, i.e., $b_-$ and $b_+$ are independent free fitting parameters.

\begin{figure*}[!]
\includegraphics[width=2.00\columnwidth]{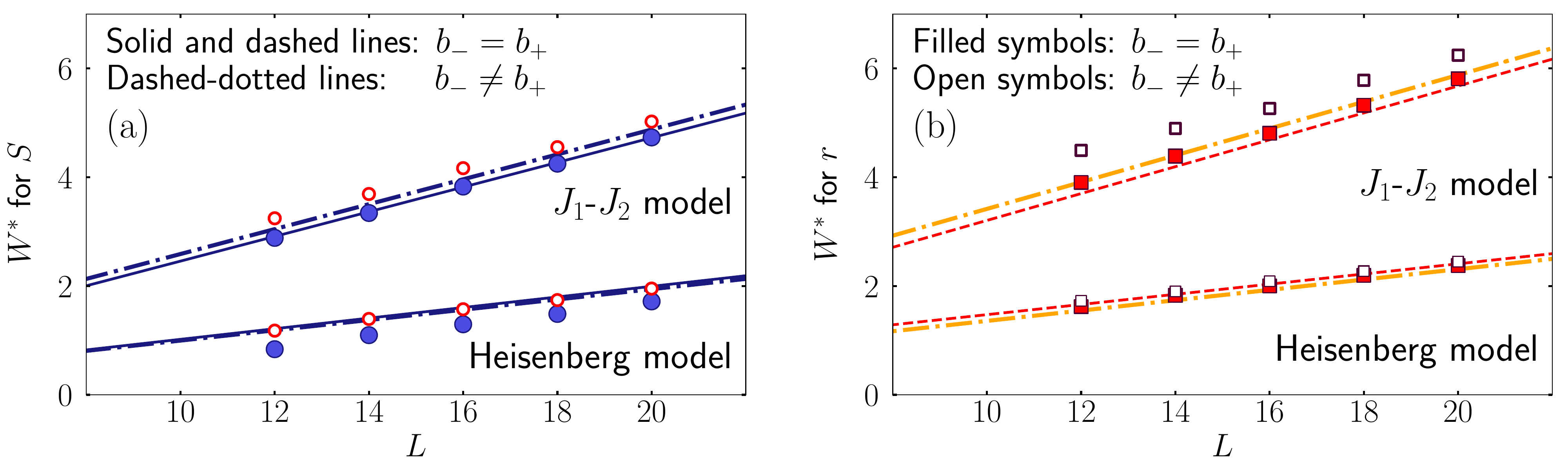}
\caption{
Ergodicity breaking transition point $W^*$ as a function of system size $L$, for the $J_1$-$J_2$ model (upper part) and the Heisenberg model (lower part).
Results for $W^*$ are obtained from the best data collapse using a KT correlation length $\xi_{\rm KT}$ from Eq.~(\ref{def_xi_kt}).
Solid and dashed lines, and filled symbols, are identical to the ones in Fig.~\ref{fig_Wcritical} of the main text (i.e., assuming $b_- = b_+$ in $\xi_{\rm KT}$).
Dashed-dotted lines and open symbols are obtained by taking independent parameters $b_- \neq b_+$.
All lines are results for a transition point ansatz $W^* = w_0 + w_1 L$ (with free parameters $w_0$ and $w_1$),
all symbols are results for a function-independent transition point $W^* = w^*(L)$ [with five free parameters for five different systems sizes $L=12,14,16,18,20$].
Results are shown for the eigenstate entanglement entropy $S$ in panel (a) and the level spacing ratio $r$ in panel (b).
}
\label{fig_Wcritical_bmbp}
\end{figure*}

We focus on functional forms of the critical disorder $W^*$ that yield the lowest cost functions, i.e., $W^* = w_0 + w_1 L$ and $W^* = w^*(L)$ [see Tables~\ref{tab:table1} and~\ref{tab:table2} in the main text].
Results for the cost functions are listed in Table~\ref{tab:table3} for the $J_1$-$J_2$ model and in Table~\ref{tab:table4} for the Heisenberg model.
The first prominent feature is that the parameter $b_+$ remains very close to the value of $b$, while the parameter $b_-$ may strongly depart from this value.
This is related to the property of the ergodicity indicators $S$ and $r$ being nearly a constant below the transition, and hence being less sensitive to the choice of $b_-$ (which then appears as quite irrelevant parameter).
The second prominent feature is that the values of parameters $w_0$, $w_1$ and $w^*(L)$ in the functional forms of the critical disorder $W^*$ remain essentially unchanged.
This is shown in Fig.~\ref{fig_Wcritical_bmbp}, where the results from both scenarios $b_- = b_+$ and $b_- \neq b_+$ exhibit fairly good agreement.
Hence, we expect that our main results remain robust against the choice of relation between $b_-$ and $b_+$.

Finally, we comment on the values of the parameters $b$ (or $b_-$, $b_+$) in the KT correlation length $\xi_{\rm KT}$ for different ergodicity indicators within the same model, listed in Tables~\ref{tab:table3} and~\ref{tab:table4}.
These values may suggest that $b$ (or $b_-$, $b_+$) are not identical for $S$ and $r$ within the same model.
Nevertheless, we refrain from making any speculations about their asymptotic values.

\begin{table}[h]
\caption{\label{tab:table3}
Cost function ${\cal C}_X$, see Eq.~(\ref{def_costfun}), and the parameters $b_-$, $b_+$ of the correlation length $\xi_{\rm KT}$, see Eq.~(\ref{def_xi_kt}), in the $J_1$-$J_2$ model.
Values of ${\cal C}_X$ are shown for two ergodicity indicators $X \in \{ S, r\}$ and for different scenarios $b_- = b_+$ and $b_- \neq b_+$.
Columns denote different functional forms of $W^*$ used in $\xi_{\rm KT}$.
Results are obtained using the data points in Figs.~\ref{fig_SvN}(a) and~\ref{fig_r}(a) for $W>0.5$.
The corresponding values of $W^*$ are shown in Fig.~\ref{fig_Wcritical_bmbp}.
}
\begin{ruledtabular}
\begin{tabular}{ l | l l }
&
$W^* = w_0 + w_1 L$ & $W^* = w^*(L)$\\
\colrule
Case $b = b_- = b_+ $: && \\
${\cal C}_S[\xi_{\rm KT}]$, $b$ & 0.34, 4.87 & 0.29, 4.90 \\
\colrule
Case $b_- \neq b_+$: && \\
${\cal C}_S[\xi_{\rm KT}], b_-, b_+$ & 0.28, 10.63, 4.75 & 0.22, 9.85, 4.53 \\
\colrule
Case $b = b_- = b_+ $: && \\
${\cal C}_r[\xi_{\rm KT}]$, $b$ & 1.01, 3.07 & 0.92, 3.00 \\
\colrule
Case $b_- \neq b_+$: && \\
${\cal C}_r[\xi_{\rm KT}], b_-, b_+$ & 0.89, 7.23, 2.84 & 0.82, 7.56, 2.54 \\
\end{tabular}
\end{ruledtabular}
\end{table}


\begin{table}[]
\caption{\label{tab:table4}
Cost function ${\cal C}_X$ for the Heisenberg model using the data points in Figs.~\ref{fig_SvN}(b) and~\ref{fig_r}(b).
Parameters are the same as in Table~\ref{tab:table3}.
}
\begin{ruledtabular}
\begin{tabular}{ l | l l }
&
$W^* = w_0 + w_1 L$ & $W^* = w^*(L)$\\
\colrule
Case $b = b_- = b_+ $: && \\
${\cal C}_S[\xi_{\rm KT}]$, $b$ & 0.46, 3.21 & 0.29, 3.59 \\
\colrule
Case $b_- \neq b_+$: && \\
${\cal C}_S[\xi_{\rm KT}], b_-, b_+$ & 0.37, 1.04, 3.34 & 0.22, 1.14, 3.35 \\
\colrule
Case $b = b_- = b_+ $: && \\
${\cal C}_r[\xi_{\rm KT}]$, $b$ & 0.51, 1.96 & 0.46, 2.04 \\
\colrule
Case $b_- \neq b_+$: && \\
${\cal C}_r[\xi_{\rm KT}], b_-, b_+$ & 0.49, 0.68, 2.23 & 0.45, 2.05, 1.95 \\
\end{tabular}
\end{ruledtabular}
\end{table}


\end{document}